\documentclass[twocolumn]{article}

\usepackage{graphicx}

\newcommand{\ie}{\textit{i.e.}}
\newcommand{\eg}{\textit{e.g.}}
\newcommand{\be}{\begin{equation}}
\newcommand{\ee}{\end{equation}}
\newcommand{\nn}{\nonumber \\}
\newcommand{\de}{\stackrel{\mbox{\tiny def}}{=}}

\newcommand{\rmd}{{\mathrm d}}

\newcommand{\blackscholes}[1]{#1^{\rm bs}}

\begin{document}

\title{A Subjective and Probabilistic Approach to Derivatives}
\author{U. Kirchner \\ ICAP \\ PO Box 1210, Houghton, 2041, South Africa \\ulrich.kirchner@icap.co.za}

\maketitle

\begin{abstract}
We propose a probabilistic framework for pricing derivatives,
which acknowledges that information and beliefs are subjective. Market prices
can be translated into implied probabilities.
In particular, futures imply returns for these implied probability distributions.
We argue that volatility is not risk, but uncertainty.
Non-normal distributions combine the risk in the left tail
with the opportunities in the right tail --- unifying the ``risk premium'' with the possible loss.
Risk and reward must be part of the same picture and expected returns must include possible losses due to risks.  

We reinterpret the Black-Scholes pricing formulas as prices for maximum-entropy probability distributions, illuminating their importance
from a new angle.

Using these ideas we show how derivatives can be priced under ``uncertain uncertainty''
and how this creates a skew for the implied volatilities.

We argue that the current standard approach based on stochastic modelling and risk-neutral pricing fails to
account for subjectivity in markets and mistreats uncertainty as risk. Furthermore,
it is founded on a questionable argument --- that uncertainty is eliminated at all cost.
\end{abstract}

\section{Introduction}

There are two ways to justify the Black-Scholes equation. On the one hand there is the standard stochastic
approach with dynamic hedging.
On the other hand one can simply calculate the expected pay-off for the log-normal probability distribution with an
expected return equal to the risk-free rate.

The second approach is considerably simpler, neater, and a direct consequence of probabilistic reasoning.
However, it cannot explain why the expected return
should be equal to the risk-free rate --- and presumably that is the reason why this derivation is hardly ever
presented.

We want to offer an alternative framework inspired by the second approach, which, in essence,
accepts the subjectivity of information/beliefs with a Bayesian interpretation of probabilities \cite{jaynes} and
reinterprets the meaning of asset prices, futures, and risk.
It yields the Black-Scholes pricing equations (but without the usual assumptions like no trading costs) as
maximum-entropy solutions for known first two moments, but the concept extends beyond
this special case.
We remove hedging arguments and replace them with probabilistic reasoning\footnote{Given
the solution hedging prescriptions can then be reconstructed.}.

One could best describe our approach as a probabilistic one embracing subjectivity and Bayesian inference, while
the traditional approach is based on statistical/stochastic ideas and methods.
These superficially appear as if they are objective in nature as they do not use prior distributions and
do not explicitly involve subjective information.
This, however, hides the fact that particular priors (usually assumptions about normality) are already assumed and hence
limit the applicability of many methods.
Also, one could argue that the neglect of available information besides historic data is rather a short-coming than an advantage.
A good overview of these issues is given in \cite{jaynes}.

This paper is organized as follows.
We start by exposing short-comings of the traditional approach in section \ref{sec-10}.
In section \ref{sec-15} we present our approach and discuss the interpretation of futures and risk. 
In section \ref{sec-20} we discuss the interpretation of Black-Scholes pricing in our context.
In section \ref{sec-25} we present applications of the probabilistic framework to
pricing under an uncertain
second moment (uncertain uncertainty), which naturally leads to an implied volatility skew,
derivative exposure and risk management.

\section{The Traditional Approach and its Problems}
\label{sec-10}

The standard approach to derivatives pricing based on stochastic models and risk-neutral measures
(see appendix \ref{a-05} for a short review)
raises a number of issues.
Firstly, it does not allow for subjectivity
as it assumes the existence of an objective ``real distribution'',
and hence does not take into account that people
have different information and believes\footnote{
A good pricing theory should be able to value an asset even for an inside-trader.
}.
This effectively leads to the idea that there is a ``correct'' value for a
derivative, independently of its current trading price --- it would possibly be ``mispriced''.

We think that this is misleading. Every market participant will assign a different value according to his state
of knowledge and beliefs. If there are sellers and buyers with different {\em subjective} valuations, it might result
in trades. These in turn might result in observed market prices. One should note that the market price then
does not necessarily represent the {\em subjective} valuation of neither the buyer nor the seller\footnote{See appendix \ref{a-20}.}.

Secondly, even if we assume this awkward ``underlying real distribution'',
we find it difficult to follow the risk-neutral hedging argument, as the
eliminated term does not only contain uncertainty and risk.
Consider for example the standard log-normal case with
\be
\rmd S = \mu S \rmd t + \sigma S \rmd X,
\ee
where $X$ is a ``normally distributed random variable''\footnote{It is questionable whether
the concept of a random variable can be consistently defined \cite{jaynes}.
One can argue that ``randomness'' is a result
of insufficient information about the state of the system. A random variable results then from uncontrolled (and hence unknown) initial conditions.}.
It is clear that $\rmd S$ becomes `increasingly deterministic' as $\sigma$ goes to zero --- for $\sigma=0$
it just expresses the deterministic drift.
Hence we want to argue that $\rmd S$ is really a mixture of a certain and uncertain component.
As one cannot eliminate the
dependence of the portfolio value change on the
different components of $\rmd S$ separately, it is not possible to follow
a hedging strategy which keeps the drift, but eliminates the randomness.

Now, by eliminating $\rmd S$ one also discards the non-random drift term and the resulting hedging is
generally (except for a drift equal to the risk free rate) suboptimal. This is most easily illustrated
in extreme (hypothetical) cases with huge (compared to the volatility) positive or negative drift.

It appears as if it is often not understood what it really {\em means} to have a drift in the distribution.
This is not just something observed in historical data, but something we claim to {\em know} about the
possibilities in the future. If we use a log-normal distribution with a 1000\% drift rate and 10\% volatility
then we also say that it is practically impossible to observe a draw-down over the next year (and even to see a return of less
than 900\%) --- and we claim to {\em know} this.

\begin{figure}
\begin{center}
\includegraphics[width=3.2in]{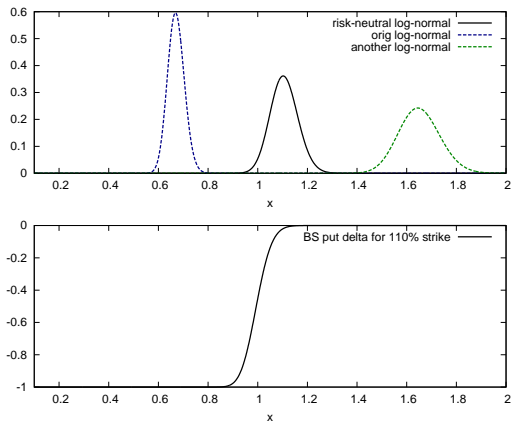}
\end{center}
\caption{
Three log-normal probability distributions (expressing our subjective state of knowledge) with different drift rates and the corresponding Black-Scholes delta for a put with
strike at the risk-free rate.
Risk-neutral hedging suggests we use the same delta independently of the drift, \ie for all three cases.
}
\label{fig-05}
\end{figure}

Yet risk-neutral pricing tells us to ignore the {\em knowledge} of the extreme drift and to hedge as if there is
a 50/50 chance of a draw-dawn (below the risk-free rate).
Having a short put-delta position will lead us with near certainty to buy back the asset at higher prices.

Figure \ref{fig-05} illustrates this. The $x$-axis shows the underlying asset value, with $x=1$ being the current spot price.
The solid line in the upper graph is a log-normal distribution with an expected return equal to the
risk-free rate (here 10\%). The lower graph shows the corresponding Black-Scholes put-delta position for a 110\% strike.
For this case it makes sense as the partial short position corresponds reasonably to the two possible outcomes.

The two coloured lines in the upper graph are two other log-normal distributions of the same (log-return) standard deviation, but with different
expected returns (drift). The blue graph represents a case for which we are practically certain that the asset price will have declined over the next year.
For such a case it is clear that one would hedge a 110\% put by immediately shorting the full amount, as we know that the asset will decline in value.

Risk-neutral pricing would tell us though to price and hedge the position as if the expected return was
actually the risk-free rate.
If the blue graph represents our state of knowledge, then this does not look like a good idea.

The case for the red line in figure \ref{fig-05} is similar. Here it is known with almost certainty that the asset value will rise
above the strike level, and hence that we should not have a short position at expiry. It is clear that in such a case we
would value the put worthless and we would also not hold a short delta position.
Again, risk-neutral pricing gives us counter-intuitive advise to ignore our {\em knowledge} of the expected return.

Note that the point is not whether we know the expected return or not.
The expected return is part of what we {\em believe}
about the asset. If we are not certain about its value we should incorporate this uncertainty using probabilistic methods,
instead of modelling our knowledge and then throwing part of the model away.

\section{An Alternative Approach}
\label{sec-15}
\paragraph{No ``real'' distribution}
In line with the Bayesian interpretation of probabilities \cite{jaynes} we do not
believe that there is an objective ``real'' distribution.
Historic data tells us what materialized, not what we should believe about the future,
and arguably not even what we should have believed (see appendix \ref{a-15}).
In fact, due to selection effects historic data can be extremely misleading --- think of hedge-fund managers
showing ``consistent alpha'' until they blow up.
It appears to be a common mistake to interpret historic data as representative of
a ``real distribution'' for future returns as suggested by statistics.

An objection often raised against the Bayesian approach is its reliance on subjective prior distributions.
We want to point out that even using historic data is a subjective matter --- we judge subjectively whether the data
can simply be extrapolated to the future.
This is fine if one is aware of above selection effects and representativity issues and judges them to be insignificant.
If, however, this treatment is due to an interpretation of the standard stochastic approach and above issues are ignored, one
makes dangerous {\em implicit} assumptions about the connection of the past and the future.
The markets have enough examples of these, like misjudged
risks on mortgages, credit default swaps, and Ponzi schemes.
In this context pricing derivatives off a ``risk-neutral'' distribution appears like a safety precaution --- too good things
(drift) should  be balanced by bad things, which are not included in the ``real distribution'' of the standard approach.

Not everything is, and can possibly be in the historic data. This data is conditional on where we are now and drawing assumptions
about the future introduces its own risks. This then has to be a subjective process involving all information available\footnote{
Interesting feedbacks arise as people use historic data as information.
If it is known how people use historic data this influences the interpretation of the historic data itself. We just want to mention here momentum effects.
} and not just historic data.  

\paragraph{Subjective and Implied Probability Distributions}
We argue that every market-participant has their own subjective probability distribution based on the information available {\em to him}.
Because people have different information available (and different levels of rationality when forming beliefs about the future) the
distributions and hence valuations are subjective.

These differing valuations then lead to trading and
the observed market prices determine implied probabilities --- one can consider them the beliefs of the market as an ``organism''\footnote{
In fact, the economy as a system seems to satisfy most criteria for a living organism. Even Darwinian evolution can be
observed in the way business structures, political systems, and commercial products change over time.
}.
Futures imply expected returns for these implied distributions, but more detailed implied distribution features can
be extracted from option prices (volatility smiles) \cite{uk}.

Hence we argue that there are only
the beliefs of each market participant (which do not have to be human, as computer trading illustrates)
and an implied probability distribution
(or for sparse data rather implied probabilities), which represents the ``beliefs of
the market''.

\paragraph{Futures}
\label{subsec-15-10}

Futures are arguably the most simple contracts and they tell us directly the implied expected return.

In the traditional approach it is argued that this is true, but it is the expected return of the
``risk-neutral'' distribution (which obviously should be the risk-free rate) and the ``real'' distribution
could have a different expected return, and in fact generally should have to allow for a risk premium.

We disagree and argue that there is nothing like a ``real'' distribution that exists objectively.

There is just an implied distribution, of which futures determine the expected return. This implied expected
return is not necessarily the risk-free rate,
but a return compatible with hedging expenses (and feasibility)
(this is a well-known fact for commodities with storage costs like oil, but also for perishable agricultural products).  
By pricing options on such assets off the futures value one implicitly uses the futures price as an expected return --- now
different to the risk-free rate.

\paragraph{Interaction of Spot and Future Markets}

We first observe here that a future arbitrage trade does not only move the futures price, but also
the spot price as assets will have to be bought or shorted. The net effect is a move towards equilibrium of
both markets. The less liquid market will adjust more than the liquid one\footnote{
Note that the spot market is not always the more liquid one. One example are illiquid grain markets.
}.

\begin{figure}
\begin{center}
\includegraphics[width=3in]{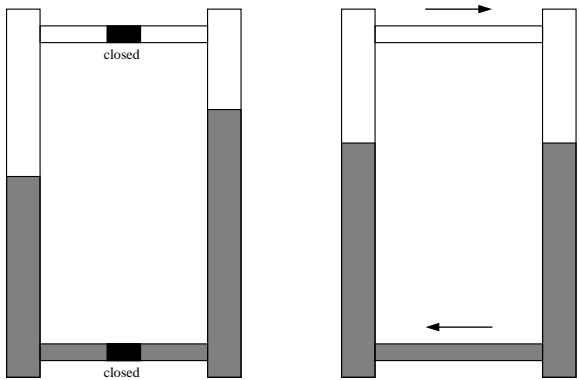}
\end{center}
\caption{An example from physics illustrating how a common equilibrium is established due to interaction.
Under the influence of gravity two columns of liquids interact. The flow of liquids is the analogue to the
flow of assets {\em through time} (\eg being removed from the spot market and stored for later disposal) between
the future and spot markets. A ``flow against time'' corresponds to short positions flowing forward in time
(like the travelling of `holes' in semiconductors). If, for example, the left side corresponds to the spot market
and the boundary between the liquids represents the asset price, then  the arrow on top represents assets being carried
to the future, and the arrow at the bottom represents borrowed cash positions to fund the arbitrage. 
The analogy could be taken further with friction and viscosity allowing slightly different levels on each side.
}
\label{fig-07}
\end{figure}

Hence it is misleading to say that arbitrage does imply futures prices. Arbitrage trades move spot {\em and}
futures markets to be compatible with hedging costs. (See figure \ref{fig-07} for an analogy from physics.)
In effect this enforces an implied expected return compatible with the hedging cost.
For assets with no storage and transaction costs and available funding at the risk-free rate, this is the risk-free rate.
But if the implied expected return is the risk-free rate, where is the risk premium? 

\subsection{Risk and Risk Premium}
\label{asdf}
Part of the confusions seems to originate from the mislabelling of uncertainty as risk.
It is clear that there are situations where we are not certain about the precise outcome, yet we are certain that
we will not have an adverse outcome. Such a situation is represented by the green line in figure \ref{fig-05} --- here
we are almost certain not to lose money (with a long position), but there is a significant uncertainty (standard deviation) about the
precise outcome.

Should uncertainty then attract a risk-premium as is for example assumed in Modern Portfolio Theory?

We would argue in two ways: Firstly, uncertainty is not what warrants a premium as it is an insufficient description of risk.
And secondly, the risk-premium
is a misleading concept as it suggests that the premium is part of the distribution, but not the risk we receive it for.

We believe risk is best defined as
the probability of an adverse outcome as suitably defined  for a certain situation\footnote{
Or through a probability distribution for adverse outcomes.
}.
Hence the risk should be part of the distribution describing our beliefs about all possible outcomes.
If we are aware of possible negative outcomes we will demand compensating possible positive outcomes.

\begin{figure}
\begin{center}
\includegraphics[width=3.2in]{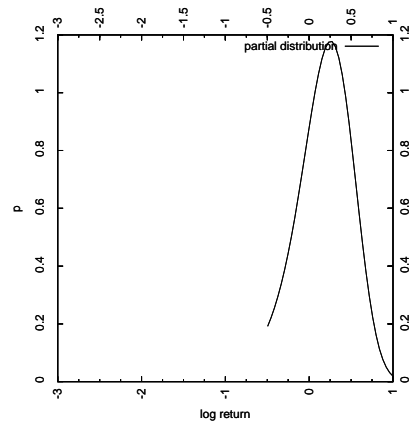}
\end{center}
\caption{
A distribution, appearing close to normal, as one might estimate from historic returns. There appears to be a mean
above the risk-free rate (10\%), which would traditionally be interpreted as a risk-premium for the uncertainty (variance or volatility).
}
\label{fig-10}
\end{figure}

\begin{figure}
\begin{center}
\includegraphics[width=3.2in]{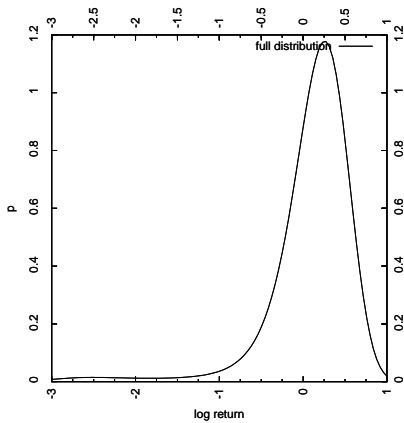}
\end{center}
\caption{
The full distribution contains the risk in the left tail. This balances with the peak above the risk-free rate to give
an expected return closer or equal to the risk-free rate.
}
\label{fig-15}
\end{figure}

Figure \ref{fig-10} shows a partial probability distribution which seems to have a significant ``risk-premium'' of about 15\% above the
risk-free rate of 10\% (all log returns).
Traditionally that would be seen as a compensation for the uncertainty (standard deviation) present.

Figure \ref{fig-15} then shows the full distribution, which is actually a maximum-entropy distribution for a mean equal to the risk-free rate
and given values for the second, third, and fourth moment. Here we see that the ``risk-premium'' is really compensated
by a fat left-hand tail. The expected value contains both --- the risk-reward, if risks do not materialize, and the
risk-penalty, if risks do materialize.

Hence we argue that ``real distributions'' (in particular log-normal with a drift higher than the risk-free rate)
are the wrong way to incorporate risk-reward, as they
incorporate a risk premium but
do not account sufficiently for the downside-risk.

\section{Connection to Classical Pricing Ideas}
\label{sec-20}
So given these concerns, why is Black-Scholes doing a reasonable\footnote{
And reasonable does not mean great. The observed skew shows that the Black-Scholes
pricing equations are insufficient. See section \ref{sec-25-10}, which addresses these issues. See \cite{haug}
for interesting comments on the history and relevance of Black-Scholes.
} job?
To answer this question let us first note a number of facts.

Firstly, one can argue that due to scale-invariance (as the actual value of a share is meaning less --- it scales with the
shares in issue) one should look at distributions over log-prices and hence log-returns.

Secondly, the log-normal probability distribution is a normal distribution for log-prices, and hence it is the
{\em maximum-entropy} distribution for {\em known} variance and mean.
This means the distribution makes the ``least assumptions'' besides what is known (the first two moments).
So whether the actual log-returns ``are normal'' or not is not the question 
 --- if we have only given the first two moments as a description of our beliefs then the normal distribution will be the right choice.

Thirdly, if we have no unique extra information (and historic data is available to everyone) then
we might want to go with the ``markets belief'' of an expected return as priced in by the futures.

Given above, the Black-Scholes pricing formulas can easily be derived as the
present value of the expected pay-off
under a log-normal probability distribution with a mean value of the risk-free rate (see appendix \ref{a-10}).
This derivation is in fact considerably simpler than the stochastic differential equation approach\footnote{It is interesting though that the Greens-function
to the Black-Scholes partial differential equation is nothing else but the log-normal distribution with the risk-free rate as mean. So
the whole effect of the detour over the stochastic pde is to eliminate the drift.} --- for example
there is no need to invoke It\^o's lemma\footnote{
The equivalent of It\^o's lemma in a probabilistic treatment is trivial: If the
probability distribution for $\ln(x)$ is a Gaussian with mean $\nu$ and standard deviation $\sigma$ then the expected value of $x$ is $e^{\nu+\sigma^2/2}$.
If the expected value of $x$ is known to be $e^{r}$ then we have to set $\nu=r-\sigma^2/2$.}.

Of course, that is not to say that we couldn't do better. Firstly, we might not be certain of the standard deviation, in which case
our result incorporates a wrong sense of certainty in a parameter --- see section \ref{sec-25-10} below, where we show that
an uncertain second moment leads to a skew.
Secondly, we might have more information about the probability distribution, \eg, higher moments.
Not using such information will of course be suboptimal.

\section{Applications}
\label{sec-25}

\subsection{Unknown Variance}
\label{sec-25-10}
Let us assume we accept the risk-free rate of return as our expected future
asset return, but we are
not certain about the second moment (variance) for a certain date.
How can we value an European option under such circumstances?

In a probabilistic framework this can be dealt with easily by marginalizing over the second moment parameter (Black-Scholes `volatility').
Starting with the definition of the derivatives value
as the present value of the expected cash-flows\footnote{
We assume here that there are no dividends and all other parameters are known, \ie expected return, time to expiry, and risk-free rate.
}
we find 
\begin{eqnarray}
V&=&e^{-rt}\int_0^\infty \rmd x \; f(x) \; p(x|I) \nn
&=&e^{-rt}\int_0^\infty \rmd \sigma \; p(\sigma|I) \int_0^\infty \rmd x \; f(x) p(x|\sigma I) \nn
&=&\int_0^\infty \rmd \sigma \; p(\sigma|I) \; \blackscholes{V}(r,\sigma),
\label{eq-25-10-5}
\end{eqnarray}
where $f(x)$ is the pay-off function, $\blackscholes{V}(r,\sigma)$ is the Black-Scholes price for given risk-free rate
and second moment $\sigma^2$ (`variance'), and $p(\sigma|I)$ is expressing our beliefs about the possible values of the (square-root of the) second moment. 

\begin{figure}
\begin{center}
\includegraphics[width=3.2in]{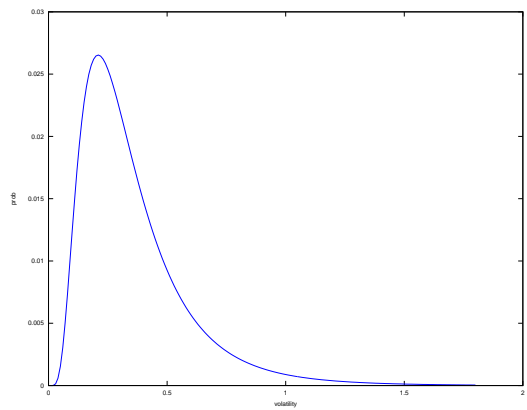}
\end{center}
\caption{
Log-normal probability distribution for the standard deviation (Black-Scholes volatility).
}
\label{fig-20}
\end{figure}

\begin{figure}
\begin{center}
\includegraphics[width=3.2in]{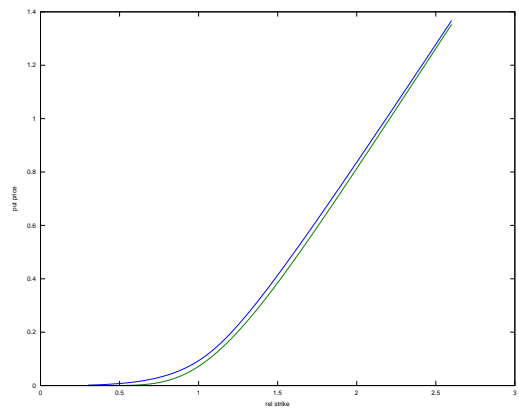}
\end{center}
\caption{
Example put prices for certain standard deviation (green line) and for a log-normal probability distribution for the standard deviation (blue line). 
}
\label{fig-25}
\end{figure}

\begin{figure}
\begin{center}
\includegraphics[width=3.2in]{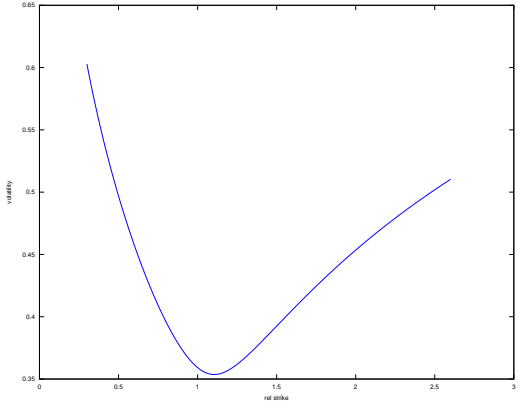}
\end{center}
\caption{
The resulting implied probability skew given the probability distribution of figure \ref{fig-20} for the second moment.
}
\label{fig-30}
\end{figure}

Figures \ref{fig-20}, \ref{fig-25}, and \ref{fig-30} give a graphical example of above relationship.
Figure \ref{fig-20} shows a probability distribution describing our beliefs about what the standard deviation (Black-Scholes `volatility') could be.
As the standard deviation is a positive quantity we consider the log-standard deviation motivated by the scale-invariance argument \cite{jaynes}.
The second order maximum entropy distribution for the standard deviation is then, just as for the price in  the Black-Scholes case,
a log-normal distribution.
The resulting put option valuations and the implied volatility skew are presented in figures \ref{fig-25} and \ref{fig-30}. 

One could now consider introducing additional ``Greeks'' for the parameters of the distribution of the standard deviation.
If, as in the example, a log-normal distribution was used then the resulting Greeks follow directly from
differentiating (\ref{eq-25-10-5}). 

\subsection{Exposure Management}
\label{sec-25-05}
Let us assume we have a set of derivatives (on the same underlying) we can invest in.
Prices of these are determined by the market, and hence are compatible with the implied probability distribution $p_m(x)$.
Our valuations and associated risks though are determined by our subjective probability distribution $p(x|I)$.

Consider a portfolio consisting of $n_i$ contracts of instrument $i$.
The current portfolio market value is
\be
\Pi^m = \sum_{i=1}^N n_i V_i^m,
\ee
where $V_i^m$ is the current market value of instrument $i$.
The value of the portfolio {\em to us} is\footnote{
Here we make the simplifying assumption that $V_i$ does not depend on the $n_j$. This is true for most
investments, but not in general. Consider the value of bottled water to you in the desert, where you are dehydrated.
The first bottle will be considerably more worth to you then the 1,000's bottle.
}
\be
\Pi = \sum_{i=1}^N n_i V_i,
\ee
where $V_i$ is the value based on our subjective probability distribution.

We suggest to maximize the objective function $\xi(n_1,\ldots,n_N)=\Pi-\Pi^m$ subject to constraints on selected risk and possibly exposure  measures (\eg no short positions).
Two examples of possible risk constraints are
\begin{itemize}
\item the probability of a loss (including initial costs) must be smaller than $y$ (this measure is only dependent on the relative number of contracts)
\item if there is a negative final portfolio value, its expected absolute value is less than $z$ (expected value
conditional on that there is a loss).
\end{itemize}
In two dimensions this optimization can easily be done graphically.

\section{Conclusion}
\label{sec-30}
We criticised the stochastic risk-neutral pricing approach for its lack of subjectivity and its mistreatment of uncertainty as risk.
In particular, risk-neutral pricing is founded on the questionable argument that any drift is worth sacrificing to eliminate
whatever uncertainty there is.

In this paper an alternative approach was presented which acknowledges that information and expectations are subjective.
Market participants should use Bayesian probabilistic reasoning to rationally express their beliefs in terms of probabilities.

Observed market prices can be translated into implied probability distributions, for which futures imply expected returns.

Given no significant extra information it is reasonable for market participants to adopt this expectation.
We showed that if a market participant is additionally sure about the standard deviation then he could rationally (due to a maximum entropy argument)
use Black-Scholes pricing to subjectively value his position. If he is not sure about the standard deviation he would
value his position according to a skew.

We commented on possible applications to derivatives exposure and risk management. 
Furthermore, we considered the case of uncertain standard deviation. For a log-normal probability distribution
for the standard deviation we illustrated how a ``volatility-smile'' was implied.

While based on simple probabilistic principles, our approach
is quite a deviation from what are current standard methods in finance
(and particularly in financial mathematics) and we believe that it offers an interesting different perspective
on financial markets.

\appendix
\section{Origins of risk-neutral Pricing}
\label{a-05}
All well known textbooks present the standard approach based on stochastic differential equations, for example \cite{hull, wilmott}.

Risk-neutral pricing has its origins in the stochastic approach to derivatives.
Hence it is based on the idea that there is a ``real'' distribution describing the
underlying return process (the ``random variable'').

In the Black-Scholes framework the ``random process'' of the underlying asset price is described by
the infinitesimal log-normal random price movements
\be
\rmd S = \mu S \rmd t + \sigma S \rmd X,
\ee
where $S$ is the asset price, $\mu$ is the drift, $\sigma$ is the volatility, and $\rmd X$ is
a normally distributed random variable.
This is the only uncertain/random contribution to the price evolution.
Hence it is argued that a portfolio is ``risk-free'' if this term is eliminated through a suitable
(dynamic) hedging strategy, in which case the portfolio should earn cash returns.  
Assuming this is done one arrives at the risk-neutral pricing formulas, which do
not depend on the drift rate $\mu$.

This concept is extended to other distributions, for which the risk-neutral measure
is linked to the ``real-world'' measure through a measure transformation, which sets the
expected return to the risk-free rate.

This then leads to the notion that there are two distributions --- the ``real'' distribution underlying the
process and the ``risk-neutral'' distribution according to which derivatives should be priced. It is important
to understand that the ``real'' distribution is thought to be necessary, because otherwise there would be
no risk-premium. Hence the dogma that futures tell us nothing about the ``real'' distribution.

\section{Deriving Black-Scholes from the log-normal Distribution}
\label{a-10}
Let $\nu$ and $\hat \sigma^2$ be the first moment and second central moment of the
log-price probability distribution for some time $t$ from now, \ie the distribution
expressing our {\em beliefs} about future market levels.
Having given no other information our best guess is the maximum-entropy distribution for the
log-price given the first two moments as constraints. This is a Gaussian distribution for the log-price, or the log-normal distribution
for the price, which is given by
\be
p(x)=\frac{1}{x \hat \sigma \sqrt{2 \pi}} \left\{ \exp{-\frac{\left[ \ln(x) -\nu \right]^2}{2 \hat \sigma^2}} \right\}.
\label{e-a5-100}
\ee
Substituting $z=\frac{\ln(x)-\nu}{\hat \sigma}$ we find
\begin{eqnarray}
\int_0^K p(x) \rmd x&=& \frac{1}{\sqrt{2 \pi}} \int_{-\infty}^{z(K)} e^{-z^2/2} \; \rmd z \nn
&=& N([\ln(K)-\nu]/\hat\sigma),
\label{e-a5-110}
\end{eqnarray}
where $N(K)$ is the cumulative normal distribution function
\be
N(x) \de \frac{1}{2 \pi} \int_{-\infty}^x \rmd z \; e^{-\frac{z^2}{2}}.
\ee

Similarly one finds with $z$ as above and the inverse $x=\exp(\hat \sigma z + \nu)$ (note
the usual completing of the square in the exponent)
\begin{eqnarray}
\int_0^K x p(x) \rmd x&=& \int_{-\infty}^{z(K)} \frac{\exp(\hat \sigma z + \nu)}{\sqrt{2 \pi}} e^{-z^2/2} \rmd z
\nn
&=& e^\frac{\nu+\hat \sigma^2}{2} \frac{1}{\sqrt{2 \pi}} \int_{-\infty}^{z(K)} e^\frac{-(z-\hat \sigma)^2}{2} \rmd z
\nn
&=& e^\frac{\nu+\hat \sigma^2}{2} N\left(\frac{\ln(K)-\nu}{\hat \sigma} -\hat \sigma\right).
\label{e-a5-120}
\end{eqnarray}
As $K \rightarrow \infty$ this gives the first moment $m_1$ of $x$ (expected value) as
\be
m_1 = e^{\nu+\hat \sigma^2/2}.
\label{e-a5-130}
\ee

Let us assume now that the first moment $m_1$ is known to be equal to the present value $x_0$ times the risk-free growth factor, \ie
\be
m_1 = x_0 e^{rt}.
\label{e-a5-140}
\ee
Substituting from above and solving for $\nu$ (the first moment of the log-price) gives
\be
\nu = \ln(x_0) + rt - \hat \sigma^2/2.
\label{e-a5-150}
\ee
For this value of $\nu$ let us define
\be
-d_2 \de [\ln(K)-\nu]/\hat \sigma= \frac{\ln\left( \frac{K}{x_0} \right) -rt +\hat\sigma^2/2}{\hat \sigma}
\label{e-a5-160}
\ee
and $d_1\de d_2+\hat \sigma$.

The value of an European put with strike $K$ is the present value of the expected cash-flow. Hence
\begin{eqnarray}
V_p&=&e^{-rt} \int_0^K \rmd x \; (K-x) p(x) 
\nn
&=& e^{-rt} \left( K N(-d_2)
- e^{\nu+\hat\sigma^2/2} N(-d_1)
\right)
\nn
&=&e^{-rt} K N(-d_2)
- x_0 N(-d_1),
\label{e-a5-170}
\end{eqnarray}
where we used (\ref{e-a5-110}) and (\ref{e-a5-120}) to evaluate the integral.
Given the second moment as an annualized variance $\hat \sigma^2=\sigma^2 t$ this agrees with the Black-Scholes pricing formula (for no dividends).

The calculation for European call and binary put/call are very similar.

\section{Reconstructing Hedging Prescriptions}

In this paper we emphasized that risk is not uncertainty.
However, if one wants to eliminate uncertainty one can derive a ``classic hedging prescription'' from a given pricing equation.
This is very similar to the usual procedure, except that the pricing has been found independently.

Let us assume we want to create a portfolio of a derivative and its underlying asset such
as to have it {\em momentarily} invariant under asset price changes.
Let $V(S)$ be our subjective value of the derivative and $\Delta$ the number of units
of the underlying asset. The portfolio value is then given by 
\be
\Pi = V + \Delta S.
\ee
Differentiating with respect to the underlying asset value $S$ and demanding that the portfolio value is invariant gives
\be
0=\frac{\partial \Pi}{\partial S} = \frac{\partial V}{\partial S} + \Delta,
\ee
and hence
\be
\Delta=-\frac{\partial V}{\partial S}.
\ee
For example, for the unknown variance case of section \ref{sec-25-10} this gives (assuming that $p(\sigma|I)$ is independent
of the spot level)
\be
\Delta=-\int_0^\infty \rmd \sigma \; p(\sigma|I) \; \blackscholes{\Delta}(r,\sigma),
\ee
where $p(\sigma|I)$ is the probability distribution for the standard deviation and $\blackscholes{\Delta}(r,\sigma)$ is
the Black-Scholes delta for given standard deviation and risk-free rate.

\section{Real Uncertainties}
\label{a-15}
One might believe that generally probabilities are about inferring a correct, but unknown value. For example,
people might ask themselves where the market will be on a certain date, and then see whether they guessed the ``correct'' value.

The problem here arises as it is implicitly assumed that the ``correct value'' exists objectively without being
dependent on the actual betting game and our associated behaviour.

Lets consider a specific example. Apparently there was a lottery system in Ireland where the ``lucky numbers'' were assigned
by a machine at the till. It so happened that the later winner was allowed by someone else to go ahead and get his numbers
first. Hence the winner claimed later that if that other person would have not let him go first, he would have not
received the winning numbers.

This certainly would be true if the numbers would have been  drawn already at that point in time (and assuming nothing
else, like a time dependent random number generator, could have affected the order of assigned numbers here).
However, what if the actual drawing process is influenced by your selection\footnote{
We argue that due to the chaotic nature of our world this is a lot more likely than the reader may think.
Some contemplation shows that there are many instances in life where a tiny change in
initial conditions has unpredictable effects. Let us only mention conception here, where
the smallest change would lead to a different genetic make-up of a living being (determining gender etc.) --- a  huge factor
in the parents further life.  
}?

In such a case the outcome is conditional on your choice. Knowing that you were right/wrong with one choice
does then not mean that you would have not been right/wrong with another choice. In other words, the lottery winner
could still have been winning (though probably with a low chance) if he would not have swapped. More importantly, the
other person who would have gotten the numbers now first probably would have not won as the outcome was conditional on them swapping.

To extract the essence consider the following: a game master preselects (lets say by some ``random method'' like
throwing a dice)
two possible outcomes (0 or 1) for A and B separately.
You are now asked to guess an answer (0 or 1).
If you say 0 the game master reveals to you the outcome drawn for A, while if you say 1 he will reveal outcome B.

Hence, having been right with your guess does not tell you whether you would have been right or wrong would
you have chosen the other answer.\footnote{
A variation on this theme is a game where a game master choses an answer based on a probability distribution
dependent on your choice. For example you choose between yes and no,
and he selects the answer to be (a) 50\% yes/50\% no if you selected yes and (b) 20\% yes/80\% no if you selected no.
Playing the game repeatedly you will now find that the distribution of outcomes depends on your bets.  
}

These situations arise if the observer is embedded in the system he is analysing and his interactions cause a meaningful disturbance in
the system. Chaotic systems are extremely sensitive to variations in initial conditions and hence one can imagine that even just the
presence of an observer (gravitational interaction) can change an outcome.

We believe that such issues are closely related to the concept of Free-Will. Knowing that your `act of prediction' is influencing the future
itself makes it impossible for you to predict the future. Hence even a deterministic system will be unpredictable {\em from within} if the
act of prediction cannot be performed without interfering with the system.

Hence for an observer inside the chaotic system the system is unpredictable and dependent on his decisions --- he perceives Free-Will.

\section{Segmented Markets and Phase Transition}
\label{a-20}
Interesting phenomena should arise if there are distinct groups of market participants, which assign
very different valuations to an asset. As long as there are enough buyers and sellers in the group attaching the highest valuation,
the situation can be stable. However, as the size of this group shrinks (for instance due to new information)
sellers might not find buyers in this group and the price can collapse. One could see this as a kind of
``phase-transition'' from one stable state into another.

As people usually subscribe to certain scenarios (``I'm a bull'' or ``I'm a bear'') and do not allow for probabilistic weightings, such  situations
appear realistic.

\end{document}